\documentclass{article}

\usepackage[backend=bibtex,url=true]{biblatex}

\usepackage{graphicx}
\usepackage{float}
\usepackage{rotating}
\usepackage{tabularx}
\usepackage{hyperref}
\usepackage[frozencache,cachedir=.]{minted}
\usepackage{xcolor}

\hypersetup{%
    colorlinks=true,
    linkcolor=black,
    citecolor=black,
    filecolor=black,
    urlcolor=black,
}

\usepackage{amsmath}
\usepackage{amssymb}
\usepackage{stmaryrd}
\usepackage{enumitem}

\usepackage{mdframed}
\BeforeBeginEnvironment{minted}{%
\begin{mdframed}[backgroundcolor=black!5,linecolor=black!1,font=\scriptsize]}
\AfterEndEnvironment{minted}{\end{mdframed}\vskip2pt}

\makeatletter
\newcommand{\ind}{\ensuremath{\rotatebox[origin=c]{90}{$\dabar@\dabar@\dabar@$}%
    \quad}}
\makeatother

\title{The process of purely event-driven programs \\ \large Revision 1}
\author{Bas van den Heuvel \\ \small%
    10343725 --- \href{mailto:vdheuvel.bas@gmail.com}{vdheuvel.bas@gmail.com}
    \\ \small %
    Supervisors: Dr.\ A. Ponse and Dr.\ ir. B. Diertens}
\date{May 25, 2018}

\begin{document}

\maketitle

\section{Introduction}

The formal specification of processes can play a crucial role in a wide range of
fields, such as factories, biological systems, railway networks, or software.
Such a formalism may help to map out the details of any process and to discover
bugs, such as deadlocks. They can also be used to validate the behaviour of a
system, or even to develop a quick prototype.

Why not apply this notion to the design of a programming language? It is more
than common that, behind the surface of its syntax, a programming language hides
the process of its execution. A programmer can discover this process by reading
the sometimes millions of lines of documentation or simply by trial-and-error.
The process is also encoded in the toolchain provided with the language:
compilers, virtual machines, interpreters, modules, etc.

In~\cite{vdHeuvel2016_pedp} I gave an abstract description of a programming
language that is concurrent by nature. By programming state machines that can
only communicate through events, one can build programs without the need of a
call stack, and with an immediate possibility to use parallel or distributed
systems.

One of the possible applications of this programming language is to provide a
tool for educational purposes. It could, for instance, be used to provide
example implementations for a distributed algorithms course. In order for this
to be useful, the language must come with tools that give insight of the
workings of such algorithms, for example in the form of Message Sequence Charts
(see~\cite{MauwReniers1994_msc}). This is another reason for formally specifying
the language's semantics; the design of such tools can be based on the
specification, instead of being based on an implementation.

By using process algebra (in particular ACP\@: the Algebra of Communicating
Processes; see~\cite{Fokkink2000_pcaintro},~\cite{BPS2001_pcahandbook}
and~\cite{Bergstra1984_acp}) combined with notions from PSF (a Process
Specification Formalism; see~\cite{MauwVeltink1993_psf}
and~\cite{Diertens2005_psf}) this paper gives a complete and thorough
description of the process of purely event-driven programs. This specification
reveals the concurrent nature of the language as described
in~\cite{vdHeuvel2016_pedp} in a formal manner.

\vfill

\newpage

\section{Purely event-driven programming}

The Purely Event-driven Programming (PEP) language is designed for
\emph{implicit concurrency}. This means that the order in which a program's
components are executed is not relevant for its result. Many programming
languages support concurrent programming, but this needs to be done explicitly
by the programmer. PEP is concurrent \emph{by design}.

Another aim of PEP is to eliminate the \emph{call stack}. Programs commonly
consist of sequences of function calls. Such calls take memory space to provide
a context, to know where to return the results to, and to remember where to
continue the program after finishing. PEP uses \emph{state machines} to get rid
of these notions, summarized in~\cite{Hohpe2006_callstack} as context,
coordination and continuation.

A state machine always has the same context, namely its own collection of
variables. One method waiting for a called method to finish is called
coordination, and remembering where to continue after a method is finished is
called continuation. Both these notions are irrelevant when two state machines
are used side-by-side. A state machine only needs to know the state it is in,
and how to react to its environment. A state never ``calls'' another state, it
merely transitions to it.

To run a state machine, one instantiates it. This way, a state machine can have
multiple running instances. Such an instance can emit events, i.e.\ a message
that has a string as type and an optional value. Other instances can be
programmed to react to certain event types, or events from specific instances.

Every state machine has a special state in which it deals with incoming events.
This state is called ``listen'', because it \emph{polls} for events, i.e.\ it
actively consults a queue to see if any events have come in. Once there is an
event, the instance's reaction settings are checked. If a reaction is set, the
instance transitions to the set state. Otherwise, the whole process starts
again.

Because of the independent nature of state machines, they are not responsible
for the actual execution of their program. The language has a central component
called Machine Control which is responsible for starting and running machines
and distributing events.

In Appendix~\ref{app:example}, accompanying a program specification, one can
find the code for a sample PEP program that simulates a very simple computer
that reads from a hard drive. However, to fully understand the PEP language
itself, please consult~\cite{vdHeuvel2016_pedp}, which contains a thorough
description of the language, together with justifications for certain design
choices as well as experiments and a Python simulation.

\section{The PEP process}

\newcommand{\then}{\ensuremath{:\shortrightarrow}}

\newcommand{\typ}[1]{\ensuremath{\text{``#1''}}}

% Event interface
\newcommand{\sndr}{\ensuremath{\text{sndr}}}
\newcommand{\dest}{\ensuremath{\text{dest}}}
\newcommand{\type}{\ensuremath{\text{type}}}
\newcommand{\ack}{\ensuremath{\text{ack}}}

% Queue functions
\newcommand{\qtop}{\ensuremath{\text{first}}}
\newcommand{\qrest}{\ensuremath{\text{tail}}}
\newcommand{\qappend}{\ensuremath{\text{append}}}

% Queue interface
\newcommand{\qempty}{\ensuremath{\text{qempty}}}
\newcommand{\deq}{\ensuremath{\text{deq}}}
\newcommand{\enq}{\ensuremath{\text{enq}}}

% Table functions
\newcommand{\qry}{\ensuremath{\text{qry}}}
\newcommand{\notfnd}{\ensuremath{\text{not\_fnd}}}
\newcommand{\fnd}{\ensuremath{\text{fnd}}}
\newcommand{\update}{\ensuremath{\text{ins}}}
\newcommand{\remove}{\ensuremath{\text{dlt}}}

% Table interface
\newcommand{\get}{\ensuremath{\text{get}}}
\newcommand{\notset}{\ensuremath{\text{not\_set}}}
\newcommand{\set}{\ensuremath{\text{set}}}
\newcommand{\unset}{\ensuremath{\text{unset}}}

\newcommand{\out}{\ensuremath{\text{out}}}

\renewcommand{\S}{\ensuremath{\mathcal{S}}}
\newcommand{\M}{\ensuremath{\mathsf{M}}}

\newcommand{\MC}{\ensuremath{\mathsf{MC}}}
\newcommand{\start}{\ensuremath{\text{start}}}
\newcommand{\newid}{\ensuremath{\text{new\_id}}}
\newcommand{\halt}{\ensuremath{\text{halt}}}
\newcommand{\distrib}{\ensuremath{\text{distrib}}}

The process of a PEP program is complex. It consists of many concurrent
subprocesses that can communicate in several ways and all have their own
specific purpose. First, I will go into some conventions used in describing the
processes. This is followed by the implementation of two important data
structures: the queue and lookup table.

Then, I give a detailed description of the process of a single running instance
of a state machine. The central part that binds a program together is Machine
Control. I finish this section by giving the processes Machine Control consists
of.

Because these processes are described as independent subprocesses, it can be
difficult to have an idea as to how the system comes together. Therefore, I have
included a schematic view of Machine Control and one state machine instance in
appendix~\ref{app:schema}.

\subsection{Conventions} \label{sec:conventions}

The following specifications are neither strictly ACP nor strictly PSF\@.
Therefore, there have to be some rules on how to name actions and how to define
communications, i.e.\ conventions. In order to make the specifications readable
for anyone, I list here the conventions used.

\begin{itemize}
    \item All actions are subscripted by an identifier. This identifier can be
        anything, from a natural number to a string, or a combination of the
        two.
    \item Actions for specific machines are subscripted with the machine's
        identifier. For example, the action for an outgoing event:
        $\out_{id}(\dots)$.
    \item If actions apply to some specific internal process, the name of the
        process is included in the identifier. For example, the action to
        dequeue an incoming event: $\deq_{in(id)}(\dots)$.
    \item Functions on internal data structures never appear outside regular
        actions. As they return immediate results on their arguments, they do
        not need to be subscripted. For example, to continue a queue after
        dequeuing: $Q_{id}(\qrest(l))$.
    \item Many processes communicate. Actions that ``send'' data after
        processing, i.e.\ outbound communication, are seen as the initiator.
        Therefore, the actions that ``receive'' data for processing, i.e.\
        inbound communication, have the same name except with a prime. Both
        sides of the communication are subscripted by the same identifier. A few
        examples:

        \begin{table}[H]
            \begin{tabularx}{\textwidth}{lllll}
                \textbf{Description} & \textbf{Outbound} & \textbf{Action} &
                    \textbf{Inbound} & \textbf{Action} \\
                Emission of an event & $\M^{out}_{id}$ & $\out_{id}(\dots)$ &
                    $\MC^{in}$ & $\out'_{id}(\dots)$ \\
                Enqueuing of an event & $\M^{prog}_{id}$ &
                    $\enq_{out(id)}(\dots)$ & $Q^{out}_{id}$ &
                    $\enq'_{out(id)}(\dots)$
            \end{tabularx}
        \end{table}

    \item Many specifications contain conditionals. They consist of some test
        that can be true or false and a question mark contained in square
        brackets, followed by a colon and an arrow. They are used to distinguish
        cases that can be determined upon entering the process term. For
        example, to check whether an event $(s, d, t, a)$ requires an
        acknowledgment: $[a = \top?] \then \dots$

    % TODO: failsafeness: only communicating actions with option to halt
    \item It is required that
        every process is designed to be \emph{fail-safe}: communicating actions
        can only be used if they are paired with the option to halt the process.
        This prevents the situation explained below.

        A process receives an event, which needs to be send to the next process.
        However, in the meantime another process has decided to halt these two
        processes. The second process receives the halt request and halts. The
        first process is now stuck with the event, without option to just quit,
        resulting in deadlock. If this process would still have the option to
        receive the halt request (i.e.\ be fail-safe), the problem would be
        solved.
\end{itemize}

\vfill

\newpage

\subsection{The event data type} \label{sec:evtype}

An event is a tuple $(s, d, t, a)$ in which \begin{itemize}[noitemsep]
    \item $s$ is a natural number identifying the sender of the event;
    \item $d$ is the identifier for its destination, or $0$ if no destination
        is given;
    \item $t$ is a string representing the event's type;
    \item $a$ is a boolean ($\top$ for true or $\bot$ for false) indicating
        whether the event requires acknowledgment.
\end{itemize}

Let $\mathbb{T}$ be the set of all event types used in the program to be
specified. We define $Ev$ to be the collection of all possible events.
\begin{align*}
    Ev := \{(s, d, t, a) \mid s \in \mathbb{N}, d \in \mathbb{N}, t \in
        \mathbb{T}, a \in \{\top, \bot\}\}
\end{align*}

% TODO: Interface as in PSF implementation, e.g. dest((s, d, t, a)) = d
Given some event $e \in Ev$, to read, for example, the sender from a tuple, one
would need to use a notation such as $e[0]$. To make the specifications more
readable, the event tuples are treated as objects. The interface to extract the
event's parameters is described in the following table.

\begin{table}[H]
    \small
    \begin{tabular}{ll}
        \textbf{Description} & \textbf{Action} \\
        Get the sender & $\sndr(e)$ \\
        Get the destination & $\dest(e)$ \\
        Get the type & $\type(e)$ \\
        Get the acknowledgment boolean & $\ack(e)$
    \end{tabular}
\end{table}

\subsection{Queues} \label{sec:queue}

A queue is the process of storing data in a fifo (first in; first out) manner.
All queues are the same, except in the type of data they store. Also, every
queue should have an identifier so other processes know which queue they are
dealing with. In this specification, all queues deal with the same type of data:
events.

A queue needs an internal list, which we call $\ell$. To initialize an empty
queue, the process can be started with the empty list $\epsilon$, as such:
$Q_{id}(\epsilon)$. The specification contains a few operations on this list
which are purely functional. I.e., they do not alter the list, but they merely
return the result of some operation. The following table describes each
operation used on some list $\ell$ with some value $e$.

% TODO: Update to how this works in PSF
\begin{table}[H]
    \small
    \begin{tabular}{ll}
        \textbf{Operation} & \textbf{Description} \\
        $\qtop(\ell)$ & Returns the first item of the list \\
        $\qrest(\ell)$ & Returns a copy of the list without the first item
            (i.e.\ without $\qtop(\ell)$) \\
        $\ell \wedge e$ & Returns a copy of the list in which $e$ is added to
            the end
    \end{tabular}
\end{table}

\vfill
\newpage

% TODO: Note on fail-safeness
A queue can be stopped by communicating a halt action with it. This particular
halt action is subscripted by $q(id)$, because the queues of instances have the
same name as their respective event handlers. Note that this specification is
fail-safe, since after every communicating action a halt request can be
received, stopping the queue.

\begin{align*}
    Q_{id}(\ell) = & [\ell = \epsilon?] \then \qempty_{id} \cdot Q_{id}(\ell)
        \\
    & + [\ell \neq \epsilon?] \then \deq_{id}(\qtop(\ell)) \cdot
        Q_{id}(\qrest(\ell)) \\
    & + \Sigma_{e \in Ev} \enq'_{id}(e) \cdot Q_{id}(\ell \wedge e) \\
    & + \halt'_{q(id)}
\end{align*}

The interface for a queue with id $id$ is depicted in the following table. Here,
$e$ is any event.

\begin{table}[H]
    \small
    \begin{tabular}{ll}
        \textbf{Description} & \textbf{Action} \\
        If the queue is empty, this action can communicate & $\qempty'_{id}$ \\
        Take the first item from the queue & $\deq'_{id}(e)$ \\
        Put an item on the queue & $\enq_{id}(e)$ \\
        Terminate the queue & $\halt_{q(id)}$
    \end{tabular}
\end{table}

\subsection{Lookup tables} \label{sec:table}

A lookup table assigns a value to a certain index. In PSF, operations on lookup
tables are purely functional. As the tables are only used by the program
process, i.e.\ they are not a shared resource, they don't need to be implemented
concurrently. However, since a state machine's program has to be able to
manipulate and access its tables in every state, it is impractical to use the
tables as arguments. Therefore, they are specified as a concurrent process. In
this way, the program process manipulates the tables by communicating actions to
it, resulting in side-effects, i.e.\ an updated table.

The variable and reaction tables differ only in the data type of their index and
their values. Therefore, in the specification we superscript the process by
collections of possible indices ($I$) and possible values ($V$). The tables
themselves use an internal lookup table with purely functional operations, which
we call $tbl$. The functionality of these operations is depicted in the
following table for some index $i \in I$ and value $v \in V$.

\begin{table}[H]
    \small
    \begin{tabularx}{\textwidth}{lX}
        \textbf{Operation} & \textbf{Description} \\
        $\qry(tbl, i)$ & Returns $\notfnd$ if the table contains no entry for
            $i$. If it does, it returns $\fnd(v)$ for some $v \in V$. \\
        $\update(tbl, i, v)$ & Returns a copy of the table with value $v$ for
            index $i$. \\
        $\remove(tbl, i)$ & Returns a copy of the table with no entry for index
            $i$.
    \end{tabularx}
\end{table}

\vfill
\newpage

% TODO: Fail-safe!
A table can be stopped by communicating a halt action with it. Again, this
specification is fail-safe, just like for the queue.

\begin{align*}
    T^{I, V}_{id}(tbl) = & \Sigma_{i \in I} ( \\
    & \ind [\qry(tbl, i) = \notfnd?] \then \notset(i) \cdot T^{I,
        V}_{id}(tbl) \\
    & \ind + \Sigma_{v \in V} ( \\
    & \ind \ind [\qry(tbl, i) = \fnd(v)?] \then \get_{id}(i, v) \cdot T^{I,
        V}_{id}(tbl) \\
    & \ind \ind + \set'_{id}(i, v) \cdot T^{I, V}_{id}(\update(tbl, i, v)) \\
    & \ind ) + \unset'_{id}(i) \cdot T^{I, V}_{id}(\remove(tbl, i)) \\
    & ) + \halt'_{id}
\end{align*}

The interface for a table with types $I$ and $V$ and id $id$ is depicted in the
following table. Here, $i$ is any index and $v$ is any value.

\begin{table}[H]
    \begin{tabular}{ll}
        \textbf{Description} & \textbf{Action} \\
        If $i: v$ is in the table, this action can communicate & $\get_{id}(i,
            v)$ \\
        If $i$ is not in the table, this action can communicate &
            $\notset_{id}(i)$ \\
        Set a value & $\set_{id}(i, v)$ \\
        Remove a value & $\unset_{id}(i)$ \\
        Terminate the table & $\halt_{id}$
    \end{tabular}
\end{table}

\subsection{State machines} \label{sec:statemach}

A PEP program actually runs by creating instances of state machines. Therefore,
we define state machines by providing templates. Such a template is subscripted
by an identifier. When an instance is made, Machine Control assigns a unique
identifier to it. This section describes how to formalise such a template for
some state machine with name ``M''.

To be able to start new instances of a machine, Machine Control needs access to
its process identifier. Therefore, we do not assign unique names for these
processes to each machine, but we use one letter $\M$ and superscript it with
the state machine's name. So, for our machine ``M'', an instance with id $id$
would be described by the process $\M^\typ{M}_{id}$.

When an instance emits an event or wants to react to events, we don't want it to
wait for Machine Control to communicate with it. Therefore, outgoing events are
put into a queue and communicated by a separate process ($\M^{out}_{id}$) and
incoming events by another ($\M^{in}_{id}$). In this way, the program
process ($\M^{prog(\typ{M})}_{id}$) can immediately continue after dealing with
events.

The identifiers of the event processes are not superscripted with the machine's
name, because they are the same for every machine. However, they do contain the
instance's id. The program process does contain the machine's name, because it
is unique to the machine.

\vfill
\newpage

This is not all. A machine needs two lookup tables: one to store how to
react to events and one to keep track of the ids of machines it started. So, the
definition of an instance includes two tables and two event queues. Let
$Vars^\typ{M}$ be the collection of all variable names used in the instance's
program.

The implicit variable name ``ctx'' is added in the specification. As it will be
provided by Machine Control, which starts the instance, the id of its context is
a parameter for the process.

The reaction table is started with the default halt reaction, except if the
context's id is 0. In this case, the machine has no context, so a halt reaction
with machine id 0 would make the machine halt if any child instance halts,
causing all machines to halt. Let $\mathbb{S}^\typ{M}$ be the set of all states
of ``M''.
\begin{align*}
    \M^\typ{M}_{id}(ctxid) =
    & \M^{prog(\typ{M})}_{id} \\
    & || T^{Vars^\typ{M} \cup \{\typ{ctx}\},
        \mathbb{N}}_{ctx(id)}([\typ{ctx}: ctxid]) \\
    & || ( \\
    & \ind [ctxid = 0?] \then T^{\{(m, t) \mid m \in \mathbb{N}, t \in
        \mathbb{T}\}, \mathbb{S}^\typ{M}}_{reactions(id)}([]) \\
    & \ind + [ctxid \neq 0?] \then T^{\{(m, t) \mid m \in \mathbb{N}, t \in
        \mathbb{T}\}, \mathbb{S}^\typ{M}}_{reactions(id)}([(ctxid, \typ{halt}):
        \typ{halt}]) \\
    & ) || Q_{in(id)}(\epsilon) || Q_{out(id)}(\epsilon) \\
    & || \M^{out}_{id} || \M^{in}_{id}
\end{align*}

\subsubsection{The program process} \label{sec:progproc}

As this project is about the process of PEP, I will not go into the details of
calculating process terms for each state. Instead, I will describe what such a
process term should look like. Let ``M'' be some state machine. For every state
$s$ of $\mathbb{S}^\typ{M}$ and every instance with id $id$ of ``M'' we define
$\S^\typ{M}_{id}$ to map state names to their corresponding processes.

An instance should always have access to the most recent event it reacted to, to
use the event's value, or to know what instance (e.g.\ the event's sender) to
return an event to. Although this behaviour is not explained in my thesis, it
was indeed used in several experiments, such as Eratosthenes' Sieve
(see~\cite[Appendix~A]{vdHeuvel2016_pedp}). Therefore, I am going to include it
in the specification.

To allow every state access to the latest event, $\S^\typ{M}_{id}$ will take it
as a second argument. This argument will only be changed in the ``listen''
state, where event reaction takes place.

Every branch (summand in process jargon) should be followed by the process of
another state. If the original code contains a branch without a state
transition, this means there is an implicit transition to the listen state. In
such cases, the branch should be changed to be followed by the listen state
process.

\vfill
\newpage

\paragraph{The listen state process} is predefined. The process first dequeues
an event and then consults its reaction table in order to decide an action.  If
the event requires an acknowledgment, an acknowledgment event is enqueued before
transitioning. If there is no event or no reaction the process tries again.

If the event queue is empty, the listen state can also take an action and then
return to the same state. It makes sure that, just as with enqueuing events
instead of communicating directly with Machine Control, an instance's program
can always perform an action.

Because there can be both a machine reaction and a regular reaction to the same
type, a priority is defined: first machine reactions, then regular ones. In this
process, this is implemented by first trying to get a machine reaction from the
table. If there is no such reaction, only the ``$\notset$'' action can
communicate. After this action, regular reactions can be tried, unless again
only ``$\notset$'' can communicate.

Because this process is rather complex, I have included a flowchart in
Appendix~\ref{app:listen}.
\begin{align*}
    \S^\typ{M}_{id}(\typ{listen}, old\_e) = & \qempty'_{in(id)} \cdot
        \S^\typ{M}_{id}(\typ{listen}, old\_e) \\
    & + \Sigma_{e \in Ev} \deq'_{in(id)}(e) \cdot ( \\
    & \ind \Sigma_{state \in \mathbb{S}^\typ{M}}
        \get_{reactions(id)}((\sndr(e), \type(e)), state) \cdot ( \\
    & \ind \ind [a = \top?] \then \enq_{out(id)}((id, s, t.\typ{ack},
        \bot)) \cdot \S^\typ{M}_{id}(state, e) \\
    & \ind \ind + [a = \bot?] \then  \S^\typ{M}_{id}(state, e) \\
    & \ind ) + \notset_{reactions(id)}((\sndr(e), \type(e))) \cdot ( \\
    & \ind \ind \Sigma_{state \in \mathbb{S}^\typ{M}}
        \get_{reactions(id)}((0, \type(e)), state) \cdot ( \\
    & \ind \ind \ind [a = \top?] \then \enq_{out(id)}((id, s, t.\typ{ack},
        \bot)) \cdot \S^\typ{M}_{id}(state, e) \\
    & \ind \ind \ind + [a = \bot?] \then \S^\typ{M}_{id}(state, e) \\
    & \ind \ind ) + \notset_{reactions(id)}((0, \type(e))) \cdot
        \S^\typ{M}_{id}(\typ{listen}, old\_e) \\
    & \ind ) \\
    & )
\end{align*}

\paragraph{The halt state process} is also predefined. If an instance is done it
should transition to this state. This state tells Machine Control to remove this
instance from its administration. Then, it tells all its internal component
processes to cease as well. This is done through communicating halt actions
subscripted by the id and the name of the process. There should be no particular
order in this, so it is done concurrently.
\begin{align*}
    \S^\typ{M}_{id}(\typ{halt}, e) = & \halt_{id} \cdot ( \\
    & \ind \halt_{ctx(id)} || \halt_{reactions(id)} \\
    & \ind || \halt_{q(in(id))} || \halt_{q(out(id))} \\
    & \ind || \halt_{in(id)} || \halt_{out(id)} \\
    & )
\end{align*}

\vfill
\newpage

\paragraph{Other state processes} can be anything, as long as every branch
continues in another state process. In this article I do not go into the details
of what is possible in a state. However, working with events is relevant, so I
have included the following table to show how to encode event actions in process
terms. The code syntax is from~\cite{vdHeuvel2016_pedp}.

\newcommand{\streep}{%
    \noindent\makebox[\linewidth]{\rule{\textwidth}{.4pt}}
}

\newcommand{\statecode}[3]{%
    \vspace{-3\parskip}
    \streep

    \begin{minipage}{.45\textwidth}
        #1
    \end{minipage}
    \hfill
    \begin{minipage}{.45\textwidth}
        \texttt{#2}
    \end{minipage}

    \begin{minipage}{.05\textwidth}
        $\Rightarrow$
    \end{minipage}
    \hfill
    \begin{minipage}{.85\textwidth}
        #3
    \end{minipage}%
}

\begin{minipage}{.45\textwidth}
    \textbf{Description}
\end{minipage}
\hfill
\begin{minipage}{.45\textwidth}
    \textbf{Code}
\end{minipage}%

\statecode{Transition to another state}{=> s}%
{$\S^\typ{M}_{id}(\typ{s}, e)$}

\statecode{Starting a machine instance}{m = ctl.start(M, ...)}%
{$\start_{id}(\typ{M}) \cdot \Sigma_{n \in \mathbb{N}} (\newid'_{id}(n)
    \cdot \set_{ctx(id)}(\typ{m}, n))$}

\statecode{Register an event reaction}{when "t" => s}%
{$\set_{reactions(id)}((0, \typ{t}), \typ{s})$}

\statecode{Unregister an event reaction}{ignore when "t"}%
{$\unset_{reactions(id)}((0, \typ{t}))$}

\statecode{Register a machine reaction}{when m emits "t" => s}%
{$\Sigma_{n \in \mathbb{N}} (\get'_{ctx(id)}(\typ{m}, n) \cdot
\set_{reactions(id)}((n, \typ{t}), \typ{s}))$}

\statecode{Unregister a machine reaction}{ignore when m emits "t"}%
{$\Sigma_{n \in \mathbb{N}} (\get'_{ctx(id)}(\typ{m}, n) \cdot
\unset_{reactions(id)}((n, \typ{t}))$}

\statecode{Emit an event}{emit ("t")}%
{$\enq_{out(id)}((id, 0, \typ{t}, \bot))$}

\statecode{Emit an event with acknowledgment}{emit ("t") => s}%
{$\enq_{out(id)}((id, 0, \typ{t}, \top)) \cdot \set_{reactions(id)}((0,
\typ{t\_ack}), \typ{s})$}

\statecode{Emit a directed event}{emit ("t") to m}%
{$\Sigma_{n \in \mathbb{N}} (\get'_{ctx(id)}(\typ{m}, n) \cdot
\enq_{out(id)}((id, n, \typ{t}, \bot)))$}

\statecode{Emit a directed event with acknowledgment}{emit ("t") to m => s}%
{$\Sigma_{n \in \mathbb{N}} (\get'_{ctx(id)}(\typ{m}, n) \cdot
\enq_{out(id)}((id, n, \typ{t}, \top)) \cdot \set_{reactions(id)}((n,
\typ{t\_ack}), \typ{s}))$}

\vfill
\newpage

\subsubsection{Event handling}\label{sec:instev}

% TODO: Argue failsafeness
The outgoing event handler takes an event from its queue and communicates it
with Machine Control. A halt action stops the process. This specification is
fail-safe, because if Machine Control's incoming event handler halts while this
process dequeues an event, it still has the option to halt.

\begin{align*}
    \M^{out}_{id} = & \Sigma_{e \in Ev} ( \\
    & \ind \deq'_{out(id)}(e) \cdot ( \\
    & \ind \ind \out_{id}(e) \cdot \M^{out}_{id} \\
    & \ind \ind + \halt'_{out(id)} \\
    & \ind ) \\
    & ) + \halt'_{out(id)}
\end{align*}

% TODO: Argue failsafeness
Incoming events are put into a queue by the incoming event handler. From there,
a machine instance's listen state dequeues the event and decides what to do with
it. A halt action stops the process. This specification is also fail-safe: if
the incoming event queue halts while this process receives an event from Machine
Control, it still has the option to halt.

\begin{align*}
    \M^{in}_{id} = & \Sigma_{e \in Ev} ( \\
    & \ind \distrib'_{id}(e) \cdot ( \\
    & \ind \ind \enq_{in(id)}(e) \cdot \M^{in}_{id} \\
    & \ind \ind + \halt'_{in(id)} \\
    & \ind ) \\
    & ) + \halt'_{in(id)}
\end{align*}

\subsection{Machine Control} \label{sec:machctl}

Machine Control is the central unit responsible for starting state machine
instances and the distribution of events. In order to keep track of the ids of
instances, it is dependent on a counter and a list. Because most events need to
be distributed to all running instances, distribution cannot be separated from
the process that keeps these ids. This part of Machine Control is the scheduler
($\MC^{sched}$).

However, the reception of events is completely independent from this. So, in
order to be able to distribute events whilst receiving events, this is done by a
separate process and a queue, as done in machine instances. This part of Machine
Control is the incoming event handler ($\MC^{in}$).

\vfill
\newpage

The process for Machine Control is defined with two parameters: $n$, a natural
number representing the amount of instances it started (i.e.\ the highest
instance id), and $\ell$, the set of ids of running instances.
\begin{align*}
    \MC(n, \ell) = & \MC^{sched}(n, \ell) \\
    & || Q_\MC(\epsilon) \\
    & || \MC^{in}
\end{align*}

\subsubsection{Scheduler}\label{sec:mcsched}

% TODO: Rewrite this text because the specification is completely different (in
% a sense)
The scheduling process can do a number of things. These things deal with the
starting and halting of instances and the distribution of events. Because
halting and starting influence the list of ids $\ell$, and events need to be
distributed to all instances in $\ell$, they cannot run concurrently. Therefore,
if the scheduler is busy halting a machine, it cannot receive a request to start
another, and vice versa.

The communication of halting and starting requests is implemented by a
\emph{recursive sum}, because it's not possible to sum over the elements of an
internal data structure (i.e.\ over the instance ids in $\ell$). The options to
start and to halt for each $id \in \ell$ are implemented in $\MC^{schedsum}$,
which recursively adds these options for the next instance id. The same holds
for the distribution of events, which is implemented by a recursive merge in
$\MC^{distrib}$.

If a request to start an instance is communicated, the scheduler communicates
the new id to the requesting instance. Then it continues its process with
updated $n$ and $\ell$, merged with the new instances' process.

If a request to halt an instance is communicated, the scheduler enqueues an
event announcing this halt. Then it checks whether all instances have halted
(i.e.\ whether its list is empty). If so, it tells its event handler and queue
to halt and then stops itself. If not, it continues with the id of the halted
instance removed from its list.

If there is an event in Machine Control's queue, a dequeue action may
communicate. After this, the scheduler proceeds to distribute the received
event. If the event has a destination, it will only be communicated with that
destination. Otherwise, the event will be concurrently communicated with all
instances with all running instances (i.e.\ the ids in $\ell$). This ensures
that all running instances receive the event and that this process will not hang
trying to communicate with non-existing instances.

\vfill
\newpage

Because this process is rather complex, I have included a flowchart in
Appendix~\ref{app:sched}.
% TODO: Rewrite these specifications
\begin{align*}
    \MC^{sched}(n, \ell) =
    & \MC^{schedsum}(n, \ell, \qrest(\ell), \qtop(\ell)) \\
    & + \Sigma_{e \in Ev} \deq'_\MC(e) \cdot ( \\
    & \ind [\dest(e) = 0?] \then ( \\
    & \ind \ind [\ell \neq (\epsilon \wedge \sndr(e))?] \then \MC^{distrib}(e,
        \qrest(\ell), \qtop(\ell)) \cdot \MC^{sched}(n, \ell) \\
    & \ind \ind + [\ell = (\epsilon \wedge \sndr(e))?] \then \MC^{sched}(n,
        \ell) \\
    & \ind ) + [\dest(e) \neq 0?] \then \distrib_{\dest(e)}(e) \cdot
        \MC^{sched}(n, \ell) \\
    & )
\end{align*}
\begin{align*}
    \MC^{schedsum}(n, orig\_\ell, \ell, id) =
    & [\ell \neq \epsilon?] \then \MC^{schedsum}(n, orig\_\ell, \qrest(\ell),
        \qtop(\ell)) \\
    & \Sigma_{M \in \mathbb{M}} \start'_{id}(M) \cdot \newid_{id}(n+1) \cdot (
        \\
    & \ind \MC^{sched}(n+1, orig\_\ell \wedge (n+1)) || \M^M_{n+1}(id) \\
    & ) + \halt'_{id} \cdot \enq_\MC((id, 0, \typ{halt}, \bot)) \cdot ( \\
    & \ind [(orig\_\ell \setminus id) = \epsilon?] \then (\halt_{\MC(in)} ||
        \halt_{q(\MC)}) \\
    & \ind + [(orig\_\ell \setminus id) \neq \epsilon?] \then \MC^{sched}(n,
        orig\_\ell \setminus id) \\
    & )
\end{align*}
\begin{align*}
    \MC^{distrib}(e, \ell, id) =
    & [\sndr(e) = id?] \then \MC^{distrib}(e, \qrest(\ell), \qtop(\ell)) \\
    & + [\sndr(e) \neq id?] \then ( \\
    & \ind [\ell = \epsilon?] \then \distrib_{id}(e) \\
    & \ind + [\ell \neq \epsilon?] \then (\distrib_{id}(e) || \MC^{distrib}(e,
        \qrest(\ell), \qtop(\ell))) \\
    & )
\end{align*}

\subsubsection{Incoming event handler}

% TODO: fail-safeness
This process is pretty simple and requires no explanation. It is fail-safe,
exactly as machine instances' event handlers are (see Section~\ref{sec:instev}).

\begin{align*}
    \MC^{in} =
    & \Sigma_{(s, d, t, a) \in Ev} ( \\
    & \ind \out'_{s}(d, t, a) \cdot ( \\
    & \ind \ind \enq_\MC((s, d, t, a)) \cdot \MC^{in} \\
    & \ind \ind + \halt'_{\MC(in)} \\
    & \ind ) \\
    & ) + \halt'_{\MC(in)}
\end{align*}

\subsection{The specification of individual programs} \label{sec:progspec}

With the information in this section, one is able to define all the process
specifications necessary to define the specification for a program. In summary,
one needs to prepare the following items:

\begin{itemize}
    % TODO: ack types + halt
    \item Define $\mathbb{T}$ to be the set of all event types present in the
        program. This defines the set of all possible events $Ev$ as in
        Section~\ref{sec:evtype}. Don't forget to include ``halt'' and the
        necessary acknowledgment event types.

    \item With $Ev$ defined, define the specification of the queue process as in
        Section~\ref{sec:queue}.

    \item Define the specification of the lookup table as in
        Section~\ref{sec:table}.

    \item Define $\mathbb{M}$ to be the set of all state machine types present
        in the program.

    \item For each state machine ``M'' of $\mathbb{M}$:

        \begin{itemize}
            % TODO: Ctx variable!
            \item Define $Vars^\typ{M}$ to be the set of all variable names used
                in the state machine. Don't forget to add ``ctx'' to this list.

            \item Define $\mathbb{S}^\typ{M}$ to be the set of all state names
                in the machine.

            \item For each state $s$ of $\mathbb{S}^\typ{M}$, define
                $\S^\typ{M}_{id}(s, e)$ to be the corresponding process
                specification, with any event $e$ as parameter, as in
                Section~\ref{sec:progproc}.

            \item Assigns the machine's initial state specification to
                $\M^{prog(\typ{M})}_{id}$.

            \item Define $\M^\typ{M}_{id}(ctxid)$ as in
                Section~\ref{sec:statemach}.
        \end{itemize}

    \item Define $\MC(n, \ell)$ as in Section~\ref{sec:machctl}.
\end{itemize}

As described in~\cite[Section~2.2.3]{vdHeuvel2016_pedp}, a program is defined
by telling Machine Control to start one specific machine. In the spirit of this
definition, the specification of a program will take as argument the initial
machine to instantiate. It starts Machine Control with one initial id of 1,
merged with the initial machine, instantiated with id 1.

% TODO: 0 instead of -1
\cite[Section~2.2.3]{vdHeuvel2016_pedp} also mentions that the context of the
first instance is an empty state machine. However, since the context of
instances in the specification is only the identifier of another instance, we
can safely choose this to be 0.

The final step is to encapsulate the program in order to force communication to
take place. We have to define the set $H$ of all actions that can only be
performed as communication.

\begin{align*}
    H := & \{\qempty_{id}, \qempty'_{id}, \deq_{id}, \deq'_{id} \mid \\
    & \quad \quad id \in \{t(n) \mid t \in \{in, out\}, n \in \mathbb{N}\} \cup
        \{\MC\}\} \\
    & \cup \{\get_{id}, \get'_{id}, \notset_{id}, \notset'_{id}, \set_{id},
        \set'_{id}, \unset_{id}, \unset'_{id} \mid \\
    & \quad \quad id \in \{t(n) \mid t \in \{ctx, reactions\}, n \in
        \mathbb{N}\}\} \\
    & \cup \{\out_{id}, \out'_{id}, \distrib_{id}, \distrib'_{id}, \start_{id},
        \start'_{id}, \newid_{id}, \newid'_{id}, \halt_{id}, \halt'_{id} \mid \\
    & \quad \quad id \in \mathbb{N}\} \\
    & \cup \{\halt_{id}, \halt'_{id} \mid \\
    & \quad \quad id \in \{q(t) \mid t \in \{in(n), out(n), \MC \mid n \in
        \mathbb{N}\}\} \\
    & \qquad \quad \cup \{ctx(n), reactions(n), in(n), out(n) \mid n \in
        \mathbb{N}\} \\
    & \qquad \quad \cup \{\MC(in)\}\}
\end{align*}

We also need to define the names of all communicating actions. As mentioned in
Section~\ref{sec:conventions}, any outbound communicating action $a$ has an
inbound counterpart $a'$. We define the communication of these actions
$\gamma(a, a')$ to be $a''$.

\begin{align*}
    \mathsf{Program}(machinetype) = \partial_H(\MC(1, \{1\}) ||
        \M^{machinetype}_1(0))
\end{align*}

To see the steps in this section in action, take a look at
Appendix~\ref{app:example}.

\section{Discussion and conclusion}

% TODO: Testing actually happened!
Firstly, although this specification \emph{seems} to give the complete picture,
it is important to test its actual workings. By implementing individual PEP
programs in PSF, it is possible to test the different aspects of the language
such as event emission and reaction, state machine instantiation, and
(cascading) halting. It also enables us to debug the design, to answer several
questions. Does the listen state priority mechanism actually work? Does the
design contain a deadlock? How does this implementation compare to the Python
simulation (see~\cite{vdHeuvel2018_pepsim})?

The example program in Appendix~\ref{app:example} has been implemented in PSF
and Go (see~\cite{golang.org}). The implementation in PSF allowed me to test
some assumptions about what is possible in these process specifications and what
is not. As an example, in an earlier version Machine Control's scheduler was
implemented with sums over the internal list of instance IDs. As mentioned in
Section~\ref{sec:mcsched}, it turned out that a different approach was required.
The implementation in Go allowed me to test the workings of the programming
language with the scheduling being out of my control, and possible parallelism
in the execution. This lead to the discovery of a possible deadlock in an
earlier design, which gave rise to the need for fail-safeness, as mentioned in
Section~\ref{sec:conventions}. A short technical report on both implementations
can be expected.

Secondly, as the attentive reader might have noticed, there is one important
aspect of PEP missing: the ability to attach values to events. Because the
language has been designed to keep one instance's memory private from other
instances, the only way to pass values is through events.

However, the notion of value passing between processes is not as simple as the
concept of it in a theoretical description of a programming language. For a
specification, all the details need to be thought of. What is a value? What data
types would be allowed, and how would these data types be implemented in a
specification? Will the amount of data be limited?

All of these important decisions need to be made before it is even possible to
specify the passing of values. Because the usage of values and variables is
necessary for programming anyway, this problem needs to be solved before
actually implementing the programming language. Therefore, it would be good to
do another project on this detail of PEP alone.

Thirdly, it would be interesting to study PEP programs using the notion of
abstraction, such as in~\cite[Chapter~5]{Fokkink2000_pcaintro}. By applying an
abstraction operator to the specification of a program, one can \emph{abstract
away} from internal operations. In this case, it could be used to compress all
actions that are concerned with events and state machines to get an
\emph{external} behaviour. This external behaviour can then be compared to the
behaviour of another implementation of the program that does not use PEP.

\vfill
\newpage

Finally, some remarks about the specification itself. During this project I
became more and more convinced that using process algebra to specify the
workings of a programming language is very useful. Working out the formulas
makes you work out every detail, revealing properties of the language that are
new to you.

On top of that, an algebraic specification can help to prevent problems during
implementation, one might discover a problem with the language's design. But
after writing a substantial amount of code, programmers usually tend to patch
the mistakes, instead of rewriting everything from scratch. A process algebra
specification reveals such errors prior to implementation, preventing the source
of a compiler or interpreter to be littered with spaghetti code.

\printbibliography
%\bibliography{refs.bib}
%\input{formalism_arXiv.bbl}

\vfill

\appendix

\begin{sidewaysfigure}
    \section{Process overview schema} \label{app:schema}
    \includegraphics[width=\textheight]{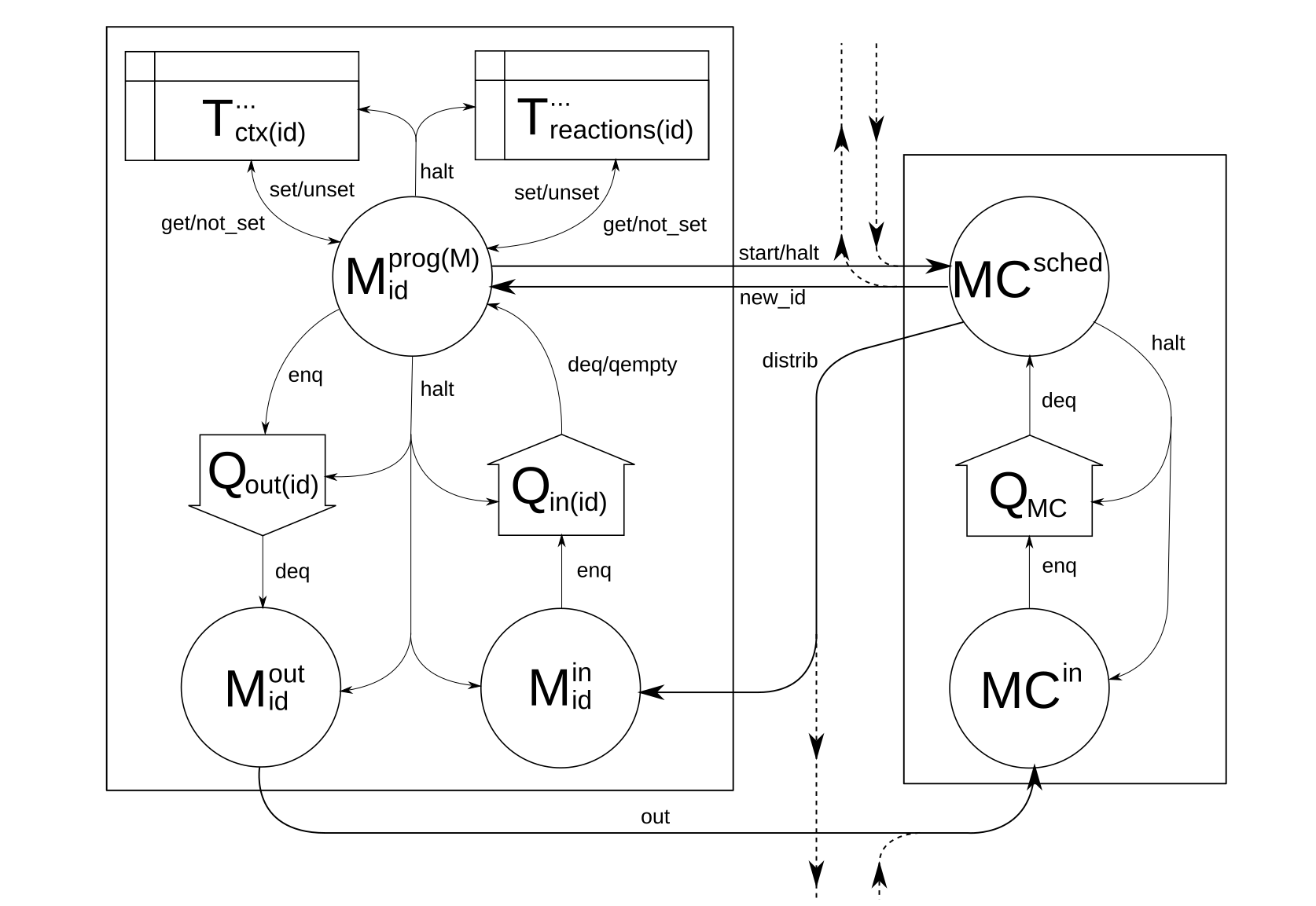}
\end{sidewaysfigure}

\section{Listen state flowchart} \label{app:listen}
\includegraphics[width=\textwidth]{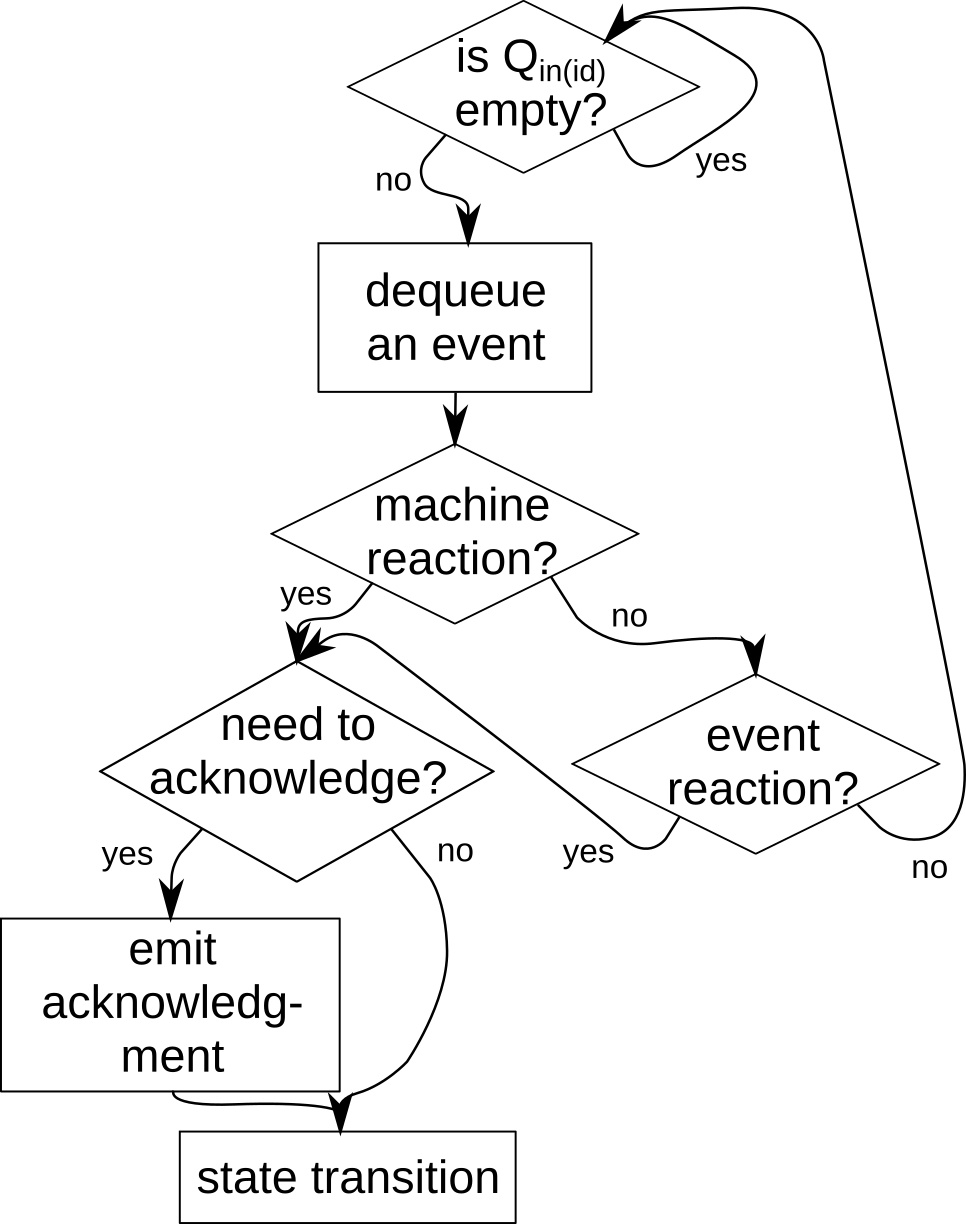}

\begin{sidewaysfigure}
    \section{Machine Control schedule state flowchart} \label{app:sched}
    \includegraphics[width=\textheight]{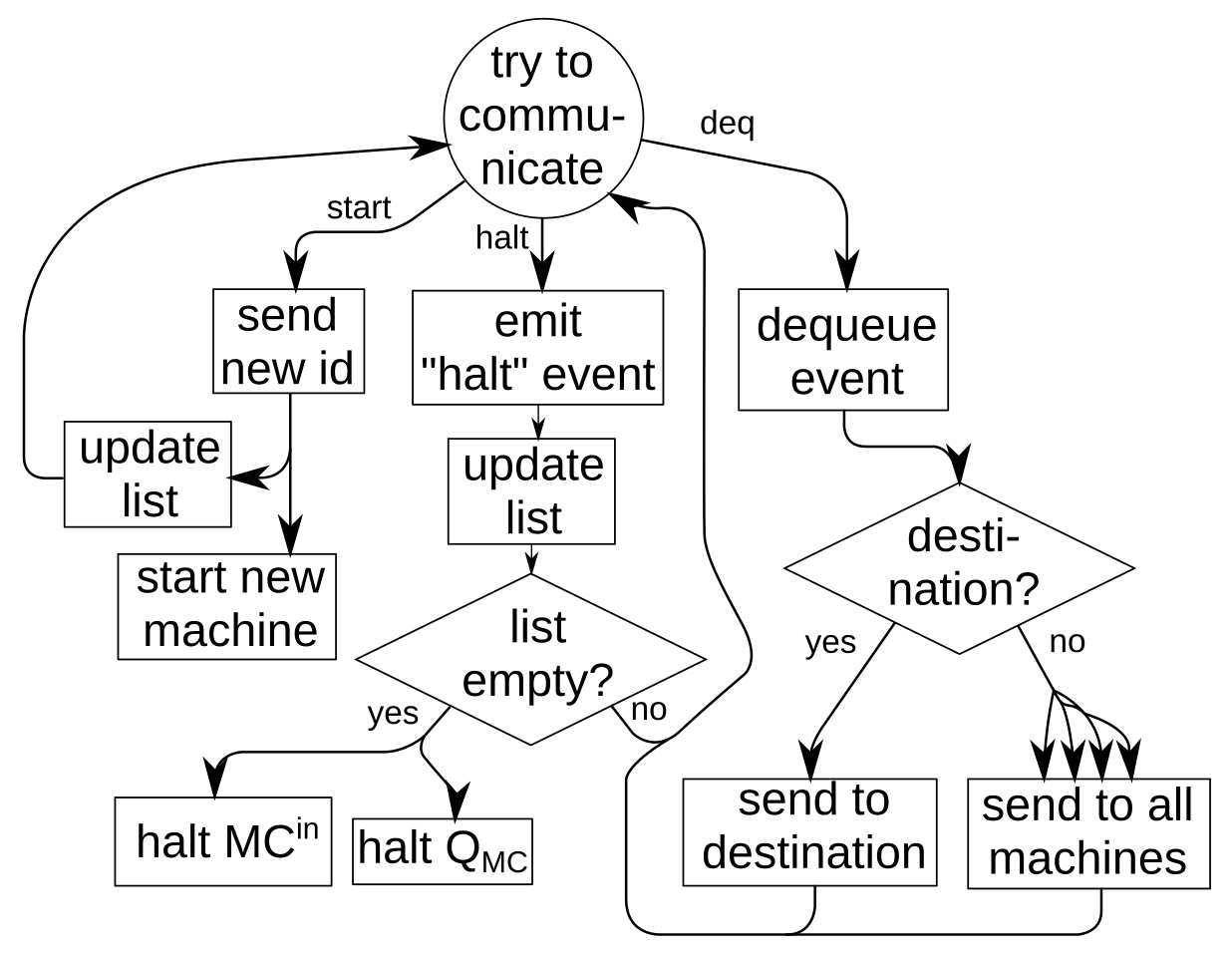}
\end{sidewaysfigure}

\section{A program specification example}
\label{app:example}

This example describes how a computer starts two programs A and B. Program B
continuously asks the CPU to give it a cycle (what for is not relevant). Program
A asks the CPU to lookup something on a hard drive. The CPU delegates this
request to the hard drive, which tells its internal head to seek the proper
data. Once found, the hard drive sends an interrupt event to the CPU\@. The CPU
then returns the found data to program A, which then shuts down the computer.

This appendix contains this program in the syntax as in~\cite{vdHeuvel2016_pedp}
(albeit with some slight alterations) and then describes how this program would
be specified according to this paper. There is also an implementation for the
Python simulator available in~\cite[\texttt{harddrive.py}]{vdHeuvel2018_pepsim}.

\subsection{Program code}

N.B.: Each state contains an argument \verb`e`, referring to the latest event.
This does not appear in~\cite{vdHeuvel2016_pedp}. This is discussed in
Section~\ref{sec:progproc}.

N.B.2: In \verb`HDHead`'s state \verb`seek` you'll notice an
\verb`or`-construct. This is also not a part of~\cite{vdHeuvel2016_pedp}, but
merely a very general way to describe non-deterministic choice. It is not really
necessary for this example to actually write code to move the arm of a hard
drive; I just want to demonstrate the possibility that seeking takes a while.

\begin{minted}[autogobble=true]{c}
    machine CPU {
        hd = null;
        prog_a = null;
        prog_b = null;
        hd_reader = null;

        init state setup(e) {
            hd = ctl.start(HD, null);
            prog_a = ctl.start(ProgA, null);
            prog_b = ctl.start(ProgB, null);

            when "cycle" => cycle;
            when "read" => hd_read;
            when "shutdown" => halt;
        }

        state cycle(e) {
            <cycle>;
        }

        state hd_read(e) {
            hd_reader = e.emitter;
            emit ("read") to hd;
            when hd emits "interrupt" => hd_interrupt;
        }

        state hd_interrupt(e) {
            emit ("return") to hd_reader;
            ignore when hd emits "interrupt";
        }
    }
\end{minted}

\vfill
\newpage

\begin{minted}[autogobble=true]{c}
    machine HD {
        hd_head = null;

        init state setup(e) {
            hd_head = ctl.start(HDHead, null);
            when ctx emits "read" => seek;
        }

        state seek(e) {
            ignore when "ctx" emits "read";
            emit ("seek") to hd_head;
            when hd_head emits "found_data" => found_data;
        }

        state found_data(e) {
            ignore when hd_head emits "found_data";
            when "read" => seek;
            emit ("interrupt") to ctx;
        }
    }

    machine HDHead {
        init state setup(e) {
            when ctx emits "seek" => seek;
        }

        state seek(e) {
            {
                emit ("found_data") to ctx;
            } or {
                => seek;
            }
        }
    }

    machine ProgA {
        init state program(e) {
            emit ("read") to ctx;
            when ctx emits "return" => finish;
        }

        state finish(e) {
            ignore when ctx emits "return";
            emit ("shutdown") to ctx;
        }
    }

    machine ProgB {
        init state cycle(e) {
            emit ("cycle") to ctx => cycle;
        }
    }

    ctl.run(CPU);
\end{minted}

\vfill

\subsection{Specification}

We follow the steps in Section~\ref{sec:progspec}. First we define the set of
all event types by going through the code and putting all used event types in a
set.
\begin{align*}
    \mathbb{T} := \{\typ{halt}, \typ{cycle}, \typ{cycle\_ack}, \typ{read},
        \typ{shutdown}, \typ{interrupt}, \typ{seek}, \typ{found\_data},
        \typ{return}\}
\end{align*}

This defines the set of all possible events $Ev$ as well as the specifications
for the queue and the lookup table, which can be directly copied. Next, we
define the set of all state machine types.
\begin{align*}
    \mathbb{M} := & \{\typ{CPU}, \typ{HD}, \typ{HDHead}, \typ{ProgA},
        \typ{ProgB}\}
\end{align*}

For each machine, we need to define the respective set of variable names and
state names. As each state machine always has a ``listen'' and a ``halt'' state,
we can define a default set of states $\mathbb{S}^{default} := \{\typ{listen},
\typ{halt}\}$. We also need to define the state process specifications. The
specifications for ``listen'' and ``halt'' are omitted because they can be
directly copied with the right machine type as superscript. We also define the
machine's program to equal its initial state (with empty event parameters).

N.B. Some specifications slightly deviate from the table in
Section~\ref{sec:progproc}, for example in $\S^\typ{CPU}_{id}(\typ{hd\_read},
e)$. This is because the same variable name is used multiple times in a
row, so the id of the relevant instance needs to be looked up just once.
\begin{align*}
    Vars^\typ{CPU} := & \{\typ{ctx}, \typ{hd}, \typ{prog\_a}, \typ{prog\_b},
        \typ{hd\_reader}\} \\
    \mathbb{S}^\typ{CPU} := & \mathbb{S}^{default} \cup \{\typ{setup},
        \typ{cycle}, \typ{hd\_read}, \typ{hd\_interrupt}\} \\
    \S^\typ{CPU}_{id}(\typ{setup}, e) = & \start_{id}(\typ{HD}) \cdot \Sigma_{n
        \in \mathbb{N}} (\newid'_{id}(n) \cdot \set_{ctx(id)}(\typ{hd}, n)) \\
    & \cdot \start_{id}(\typ{ProgA}) \cdot \Sigma_{n \in
        \mathbb{N}} (\newid'_{id}(n) \cdot \set_{ctx(id)}(\typ{prog\_a}, n)) \\
    & \cdot \start_{id}(\typ{ProgB}) \cdot \Sigma_{n \in
        \mathbb{N}} (\newid'_{id}(n) \cdot \set_{ctx(id)}(\typ{prog\_b}, n)) \\
    & \cdot \set_{reactions(id)}((0, \typ{cycle}), \typ{cycle}) \\
    & \cdot \set_{reactions(id)}((0, \typ{read}), \typ{hd\_read}) \\
    & \cdot \set_{reactions(id)}((0, \typ{shutdown}), \typ{halt}) \\
    & \cdot \S^\typ{CPU}_{id}(\typ{listen}, e) \\
    \S^\typ{CPU}_{id}(\typ{cycle}, e) = & \langle cycle \rangle \cdot
        \S^\typ{CPU}_{id}(\typ{listen}, e) \\
    \S^\typ{CPU}_{id}(\typ{hd\_read}, e) = &
        \set_{ctx(id)}(\typ{hd\_reader}, \sndr(e)) \\
    & \cdot \Sigma_{n \in \mathbb{N}} (\get'_{ctx(id)}(\typ{hd}, n) \cdot
        \enq_{out(id)}((id, n, \typ{read}, \bot)) \\
    & \ind \cdot \set_{reactions(id)}((n, \typ{interrupt}),
        \typ{hd\_interrupt})) \\
    & \cdot \S^\typ{CPU}_{id}(\typ{listen}, e) \\
    \S^\typ{CPU}_{id}(\typ{hd\_interrupt}, e) = & \Sigma_{n \in \mathbb{N}}
        (\get'_{ctx(id)}(\typ{hd\_reader}, n) \cdot \enq_{out(id)}((id, n,
        \typ{return}, \bot))) \\
    & \cdot \Sigma_{n \in \mathbb{N}} (\get'_{ctx(id)}(\typ{hd}, n) \cdot
        \unset_{reactions(id)}((n, \typ{interrupt}))) \\
    & \cdot \S^\typ{CPU}_{id}(\typ{listen}, e) \\
    \M^{prog(\typ{CPU})}_{id} = & \S^\typ{CPU}_{id}(\typ{setup}, (0, 0, \typ{},
    \bot))
\end{align*}

\begin{align*}
    Vars^\typ{HD} := & \{\typ{ctx}, \typ{hd\_head}\} \\
    \mathbb{S}^\typ{HD} := & \mathbb{S}^{default} \cup \{\typ{setup},
        \typ{seek}, \typ{found\_data}\} \\
    \S^\typ{HD}(\typ{setup}, e) = & \start_{id}(\typ{HDHead}) \cdot
        \Sigma_{n \in \mathbb{N}} (\newid'_{id}(n) \cdot
        \set_{ctx(id)}(\typ{hd\_head}, n)) \\
    & \cdot \Sigma_{n \in \mathbb{N}} (\get'_{id}(\typ{ctx}, n) \cdot
        \set_{reactions(id)}((n, \typ{read}), \typ{seek})) \\
    & \cdot \S^\typ{HD}_{id}(\typ{listen}, e) \\
    \S^\typ{HD}(\typ{seek}, e) = & \Sigma_{n \in \mathbb{N}}
        (\get'_{ctx(id)}(\typ{ctx}, n) \cdot \unset_{reactions(id)}((n,
        \typ{read}))) \\
    & \cdot \Sigma_{n \in \mathbb{N}} (\get'_{ctx(id)}(\typ{hd\_head}, n) \cdot
        \enq_{out(id)}((id, n, \typ{seek}, \bot)) \\
    & \ind \cdot \set_{reactions(id)}((n, \typ{found\_data}),
        \typ{found\_data})) \\
    & \cdot \S^\typ{HD}_{id}(\typ{listen}, e) \\
    \S^\typ{HD}(\typ{found\_data}, e) = & \Sigma_{n \in \mathbb{N}}
        (\get'_{ctx(id)}(\typ{hd\_head}, n) \cdot \unset_{reactions(id)}((n,
        \typ{found\_data}))) \\
    & \cdot \Sigma_{n \in \mathbb{N}} (\get'_{ctx(id)}(\typ{ctx}, n) \cdot
        \set_{reactions(id)}((n, \typ{read}), \typ{seek}) \\
    & \ind \cdot \enq_{out(id)}((id, n, \typ{interrupt}, \bot))) \\
    & \cdot \S^\typ{HD}_{id}(\typ{listen}, e) \\
    \M^{prog(\typ{HD})}_{id} = & \S^\typ{HD}_{id}(\typ{setup}, (0, 0, \typ{},
    \bot))
\end{align*}

\begin{align*}
    Vars^\typ{HDHead} := & \{\typ{ctx}\} \\
    \mathbb{S}^\typ{HDHead} := & \mathbb{S}^{default} \cup \{\typ{setup},
        \typ{seek}\} \\
    \S^\typ{HDHead}_{id}(\typ{setup}, e) = & \Sigma_{n \in \mathbb{N}}
        (\get'_{ctx(id)}(\typ{ctx}, n) \cdot \set_{reactions(id)}((n,
        \typ{seek}), \typ{seek})) \\
    & \cdot \S^\typ{HDHead}_{id}(\typ{listen}, e) \\
    \S^\typ{HDHead}_{id}(\typ{seek}, e) = & ( \\
    & \ind \Sigma_{n \in \mathbb{N}} (\get'_{ctx(id)}(\typ{ctx}, n) \cdot
        \enq_{out(id)}((id, n, \typ{found\_data}, \bot))) \\
    & \ind \cdot \S^\typ{HDHead}_{id}(\typ{listen}, e) \\
    & ) + ( \\
    & \ind \langle seek \rangle \cdot \S^\typ{HDHead}_{id}(\typ{seek}, e) \\
    & ) \\
    \M^{prog(\typ{HDHead})}_{id} = & \S^\typ{HDHead}_{id}(\typ{setup}, (0, 0,
        \typ{}, \bot))
\end{align*}

\begin{align*}
    Vars^\typ{ProgA} := & \{\typ{ctx}\} \\
    \mathbb{S}^\typ{ProgA} := & \mathbb{S}^{default} \cup \{\typ{program},
        \typ{finish}\} \\
    \S^\typ{ProgA}_{id}(\typ{program}, e) = & \Sigma_{n \in \mathbb{N}}
        (\get'_{ctx(id)}(\typ{ctx}, n) \cdot \enq_{out(id)}((id, n, \typ{read},
        \bot)) \\
    & \ind \cdot \set_{reactions(id)}((n, \typ{return}), \typ{finish})) \\
    & \cdot \S^\typ{ProgA}_{id}(\typ{listen}, e) \\
    \S^\typ{ProgA}_{id}(\typ{finish}, e) = & \Sigma_{n \in \mathbb{N}}
        (\get'_{ctx(id)}(\typ{ctx}, n) \cdot \unset_{reactions(id)}((n,
        \typ{return})) \\
    & \ind \cdot \enq_{out(id)}((id, n, \typ{shutdown}, \bot))) \\
    & \cdot \S^\typ{ProgA}_{id}(\typ{listen}, e) \\
    \M^{prog(\typ{ProgA})}_{id} = & \S^\typ{ProgA}_{id}(\typ{program}, (0, 0,
        \typ{}, \bot))
\end{align*}

\begin{align*}
    Vars^\typ{ProgB} := & \{\typ{ctx}\} \\
    \mathbb{S}^\typ{ProgB} := & \mathbb{S}^{default} \cup \{\typ{cycle}\} \\
    \S^\typ{ProgB}_{id}(\typ{cycle}, e) = & \Sigma_{n \in \mathbb{N}}
        (\get'_{ctx(id)}(\typ{ctx}, n) \cdot \enq_{out(id)}((id, n, \typ{cycle},
        \top)) \\
    & \ind \cdot \set_{reactions(id)}((n, \typ{cycle\_ack}), \typ{cycle})) \\
    & \S^\typ{ProgB}_{id}(\typ{listen}, e) \\
    \M^{prog(\typ{ProgB})}_{id} = & \S^\typ{ProgB}_{id}(\typ{cycle}, (0, 0,
        \typ{}, \bot))
\end{align*}

\vfill

These are the ingredients to define the instance process for each state machine.
These definitions are mostly the same, so I only give the one for ``CPU''.
\begin{align*}
    \M^\typ{CPU}_{id}(ctxid) = & (\M^{prog(\typ{CPU})}_{id} \\
    & \ind || T^{Vars^\typ{CPU} \cup \{\typ{ctx}\},
        \mathbb{N}}_{ctx(id)}([\typ{ctx}: ctxid]) \\
    & \ind || T^{\{(m, t) \mid m \in \mathbb{N}, t \in \mathbb{T}\},
        \mathbb{S}^\typ{CPU}}_{reactions(id)}([(ctxid, \typ{halt}):
        \typ{halt}])) \\
    & || Q_{in(id)}(\epsilon) || Q_{out(id)}(\epsilon) \\
    & || \M^{out}_{id} || \M^{in}_{id}
\end{align*}

Next, we copy the definition for $\MC(n, \ell)$. Finally, we define the program.
\begin{align*}
    \mathsf{harddisk} = \mathsf{Program}(\typ{CPU}) = \partial_H(\MC(1, \{1\})
        || \M^\typ{CPU}_1(0))
\end{align*}

\end{document}